\documentclass[superscriptaddress,aps,preprintnumbers,amsmath,amssymb,sort&compress,nofootinbib,10pt,twocolumn,prd, floatfix]{revtex4-2}

\usepackage{graphicx}
\usepackage[bookmarksopen,colorlinks=true,linkcolor=blue,citecolor=blue,urlcolor=blue]{hyperref}

\usepackage{color, comment, bm, ascmac, microtype, fonttable}

\newcommand{\CosmoLattice}{$\mathcal{C}$osmo$\mathcal{L}$attice}

\newcommand{\colorBigCrunchPositive}{red}
\newcommand{\colorBigCrunchNegative}{purple}
\newcommand{\colorCyclic}{sky blue}
\newcommand{\colorShortInflation}{green}
\newcommand{\colorLongInflationTachyonic}{orange}
\newcommand{\colorLongInflation}{yellow}

\newcommand{\colorChoice}{white}
\newcommand{\colorFlatness}{black}

\newcommand{\lineIdea}{thin orange}
\newcommand{\lineBenchMark}{thick blue}

\begin{document}
\preprint{YITP-23-56}
\preprint{CTPU-PTC-23-14}
\preprint{TU-1186}

\title{
Dissipative genesis of the inflationary Universe
}

\author{Hiroki Matsui}
\affiliation{Center for Gravitational Physics and Quantum Information, 
Yukawa Institute for Theoretical Physics,
Kyoto University, 606-8502, Kyoto, Japan}

\author{Alexandros Papageorgiou}
\affiliation{Particle Theory  and Cosmology Group, Center for Theoretical Physics of the Universe, 
 Institute for Basic Science (IBS), Daejeon, 34126, Korea}

\author{Fuminobu Takahashi}
\affiliation{Department of Physics, Tohoku University, Sendai, 980-8578, Miyagi, Japan}

\author{Takahiro Terada}
\affiliation{Particle Theory  and Cosmology Group, Center for Theoretical Physics of the Universe, 
Institute for Basic Science (IBS), Daejeon, 34126, Korea}

\begin{abstract}
We study an inflation model with a flat scalar potential supported by observations and find that slow-roll inflation can emerge after a quasi-cyclic phase of the Universe, where it undergoes repeated expansions and contractions for a finite time period. The initial conditions and the positive spatial curvature required for such nontrivial dynamics align with the quantum creation of the Universe. The key ingredients that trigger inflation are dissipative interactions of the inflaton, which are necessary to reheat the Universe after inflation and thus give us an observational handle on pre-inflationary physics. Our discovery implies that inflation occurs more robustly after the creation.
\end{abstract}

\date{\today}
\maketitle

\section{Introduction}

One of the fundamental questions in  nature is the origin of the Universe. The discovery of the cosmic microwave background (CMB)~\cite{Penzias:1965wn} established the big bang cosmology, whose problems (horizon, flatness, monopole puzzles) were solved in the cosmic inflation paradigm, in which the primordial Universe experiences a period of nearly exponential expansion~\cite{Starobinsky:1980te, Sato:1980yn, Guth:1980zm, Linde:1981mu, Albrecht:1982wi, Linde:1983gd}. Quantum fluctuations of the inflaton, the scalar field driving inflation,  serve as seeds for the large-scale structure (LSS) of the Universe.  Its simplest version, i.e., the single-field slow-roll inflation scenario is consistent with cosmological observations of CMB~\cite{Planck:2018vyg, Planck:2018jri} and LSS~\cite{SDSS:2023tbz}. Alternatives to inflation include bounce cosmology, in which a contracting Universe transitions into an expansion phase, and cyclic cosmology,\footnote{
We consider the cyclicity of the scale factor rather than the Hubble parameter~\cite{Ijjas:2019pyf}.
} in which the Universe undergoes repeated expansion and unlimited contraction (see reviews~\cite{Novello:2008ra, Battefeld:2014uga, Brandenberger:2016vhg}). A more ambitious idea to understand the beginning, if it exists, of the Universe is quantum creation of the Universe from ``nothing'' including the tunneling proposals~\cite{Linde:1983cm, Linde:1983mx, Linde:1984ir, Rubakov:1984bh, Zeldovich:1984vk, Vilenkin:1984wp, Vilenkin:1986cy, Vilenkin:1987kf} and the no-boundary proposal~\cite{Hawking:1981gb, Hartle:1983ai, Hawking:1983hj}.  In this letter, we point out that these diverse ideas coherently merge into a new consistent picture of the Universe. 

In fact, none of the above individually provide us a fully satisfactory cosmology.  Inflation does not necessarily resolve the singularity of the big bang~\cite{Borde:2001nh}.\footnote{
Past-incompleteness shown in Ref.~\cite{Borde:2001nh} does not necessarily imply physical singularity~\cite{Yoshida:2018ndv, Nomura:2021lzz, Lesnefsky:2022fen, Geshnizjani:2023hyd}.
}  
Bounce cosmology assumes an infinitely large homogeneous contracting universe in the infinite past, which is possible but not entirely satisfactory, at least to us. The cyclic Universe is a beautiful idea, but generally conflicts with the second law of thermodynamics and is unstable (see, e.g., Ref.~\cite{Tolman:1931fei}).  For the quantum creation of the Universe to be predictive, we need to better understand the classical dynamics that occur after creation.  For instance, sufficiently long inflation immediately after creation is not necessarily guaranteed, so we need to explore whether and how our Universe can arise in such cases. 

Since the beginning of the Universe is a highly speculative topic with no direct observational evidence, it will be useful as a first step to incorporate our knowledge about CMB observables, which suggest inflation models with a relatively flat potential~\cite{Planck:2018jri}.  We study the classical dynamics of the Universe with such a flat scalar potential and with the initial conditions and the positive spatial curvature motivated by the quantum creation of the Universe. In our previous work with such a setup, we found homogeneous solutions representing a cyclic or ``vibrating'' Universe when the initial inflaton field value is around the edge of the flat part of the potential~\cite{Matsui:2019ygj}.  This is the starting point from which creation from nothing, cyclic cosmology, and inflation will eventually unite.

The first step is to notice the possibility that inflation might occur because of the instability of the quasi-cyclic epoch~\cite{Barrow:1995cfa, Mulryne:2005ef, Biswas:2008kj, Biswas:2009fv, Barrow:2017yqt, Yoshida:2019pgn}. 
In our setup, a plateau-type potential necessarily has a field region in which tachyonic instability occurs.  Also, if we assume that inflation occurs after the quasi-cyclic epoch in some way, the inflaton must dissipate into other fields and (re)heat the Universe. Such interactions inevitably participate in the instability process of the quasi-cyclic epoch. The purpose of this letter is to study this instability process and whether and how an inflation phase emerges.  We find that dissipative interactions can turn the quasi-cyclic Universe into the inflationary Universe before the tachyonic instability significantly develops for appropriate parameter values.

We do \emph{not} introduce modifications of gravity beyond General Relativity, exotic matter violating the null energy condition, a negative cosmological constant, negative Casimir energy, or a singular bounce.  Also, our mechanism is based on classical dynamics of the inflaton with sub-Planckian energy density, and it does not involve quantum tunneling from the quasi-cyclic period into inflation.

\section{Setup}
Throughout this letter, we use the natural unit $c = \hbar = 8\pi G = 1$. 
Our action is given by the following Lagrangian density:
\begin{align}
    \mathcal{L} = \frac{1}{2}R - \frac{1}{2}g^{\mu\nu}\partial_\mu \phi \partial_\nu \phi - V(\phi) + \mathcal{L}_\text{int} + \mathcal{L}_\text{matter}, \label{full_theory}
\end{align}
where $R$ is the Ricci scalar, $g^{\mu\nu}$ the inverse metric, $\phi$ the scalar field we are interested in, $V(\phi)$ its potential, $\mathcal{L}_\text{int}$ the interaction between $\phi$ and other fields, and $\mathcal{L}_\text{matter}$ the Lagrangian density for the other fields.  As a concrete example, we consider the potential~\cite{Ferrara:2013rsa} 
\begin{align}
    V(\phi) = V_0 \tanh^2 \left(\frac{\phi}{\phi_0} \right), \label{V}
\end{align}
where $V_0$ is the overall scale of the potential and $\phi_0$ characterizes the width of the valley of the potential. This type of potential can be obtained, e.g., after switching to the Einstein frame and/or canonical normalization if the original model possesses an approximate scale invariance~\cite{Starobinsky:1980te, Bezrukov:2007ep, Kallosh:2013yoa, Galante:2014ifa}. The shift symmetry of $\phi$ can be protected from quantum corrections~\cite{Bezrukov:2010jz}.  If inflation successfully occurs, this potential leads to inflationary observables consistent with the CMB data such as
\begin{align}
    A_\text{s} = & \frac{N_\mathrm{e}^2 V_0}{3 \pi^2 \phi_0^2}, &
    n_\text{s} = & 1 - \frac{2}{N_\mathrm{e}}, &
    r =& \frac{2 \phi_0^2}{N_\mathrm{e}^2}, 
\end{align}
where $A_\text{s}$, $n_\text{s}$, $r$, and $N_\mathrm{e}$ are the amplitude of the primordial curvature perturbations, the spectral index, the tensor-to-scalar ratio, and the e-folding number between the horizon exit of the CMB pivot scale and the end of inflation. 

We assume homogeneity and isotropy of the Universe and take the Friedmann-Lema\^{i}tre-Robertson-Walker ansatz for the metric
\begin{align}
    \mathrm{d}s^2 = - \mathrm{d}t^2 + a(t)^2 \left( \frac{\mathrm{d}r^2}{1- K r^2} + r^2 \mathrm{d} \Omega_2^2  \right),
\end{align}
where $a(t)$ is the scale factor of the Universe, $K(>0)$ is the positive spatial curvature representing the closed topology of the Universe created from nothing, and $\mathrm{d}\Omega_2^2$ is the line element for the two-dimensional unit sphere. 

We adopt the initial condition motivated by the quantum creation of the Universe~\cite{Vilenkin:1987kf}
\begin{align}
    a(0) =& \sqrt{\frac{3 K}{V(\phi(0))}}, &
    \dot{a}(0) =& 0, &
    \dot{\phi}(0) =& 0,
\end{align}
with the homogeneous field value $\phi(0)$ satisfying $|V'| \lesssim V$, where a dot and a prime denote a derivative with respect to time $t$ and $\phi$, respectively.  $K$ can be fixed to an arbitrary positive value by rescaling $a$.

When the scalar field is initially on the plateau, sufficiently long inflation is possible. In this letter, we are interested in more subtle cases in which the scalar field is initially around the edge of the plateau. In this case, there are cyclic solutions at the homogeneous level~\cite{Matsui:2019ygj}. One reason why this simple setup with a single scalar field leads to cyclic solutions is that its field-dependent equation-of-state parameter $w = P/\rho$ with $P$ and $\rho$ being pressure and energy density gives rise to a feedback mechanism~\cite{Matsui:2019ygj}. 

\section{Emergence of Inflation}

To reheat the Universe, the inflaton must dissipate its energy into other, possibly light and relativistic, degrees of freedom.  Details of this depend on $\mathcal{L}_\text{int}$ and $\mathcal{L}_\text{matter}$, but we model them by the following Friedmann and Boltzmann equations:
\begin{align}
    &\dot{H} = - \frac{1}{2}\dot{\phi}^2 - \frac{2}{3}\rho_\text{r} + \frac{K}{a^2}, \\
    &\ddot{\phi} + \left(3H + \Gamma \right) \dot{\phi} + V' = 0, \\
    &\dot{\rho}_\text{r} + 4 H \rho_\text{r} = \Gamma \dot{\phi}^2,
\end{align}
where $H = \dot{a}/a$ is the Hubble expansion/contraction rate, $\Gamma$ the dissipation rate, and $\rho_\text{r}$ the energy density of radiation, i.e., relativistic degrees of freedom. 
We will use the above simplified equations in our analysis, since our scenario does not rely on the details of the equations of motion, but note that it is not always possible to derive this form from the underlying Lagrangian.  
See, e.g., Refs.~\cite{Morikawa:1984dz, Morikawa:1986rp, Gleiser:1993ea, Berera:1998gx}.  The dissipation rate $\Gamma$ may depend on the field value and the temperature if a thermal bath is populated.  The dissipation channel may include decay~\cite{Abbott:1982hn, 
 Kofman:1997yn}, scattering~\cite{Moroi:2014mqa}, thermal friction due to sphaleron transitions~\cite{McLerran:1990de,Moore:2010jd,Bodeker:1999gx,Arnold:1999ux,Arnold:1999uy,Moore:2000mx,Moore:2000ara}, etc. 
Interesting multiple cosmological roles and underlying quantum field theoretic origin of $\Gamma$ were discussed extensively in the context of warm inflation~\cite{Berera:1995ie, Berera:1998gx, Berera:2000xz, Bartrum:2014fla, Bastero-Gil:2016mrl}. We do not restrict the inflation that emerges (to be shown below) to be warm or cold. It is possible to realize cold inflation~\cite{Matsui:2023ezh}.

To illustrate essential points of our mechanism, we consider a toy model with a constant $\Gamma$, whereas we study a more realistic model in our companion paper~\cite{Matsui:2023ezh}, where we delineate possible ultraviolet completion of the models to explain a rather large coupling required for the mechanism to work.

\begin{figure}[htbp]
    \centering
    \includegraphics[width=0.99\columnwidth]{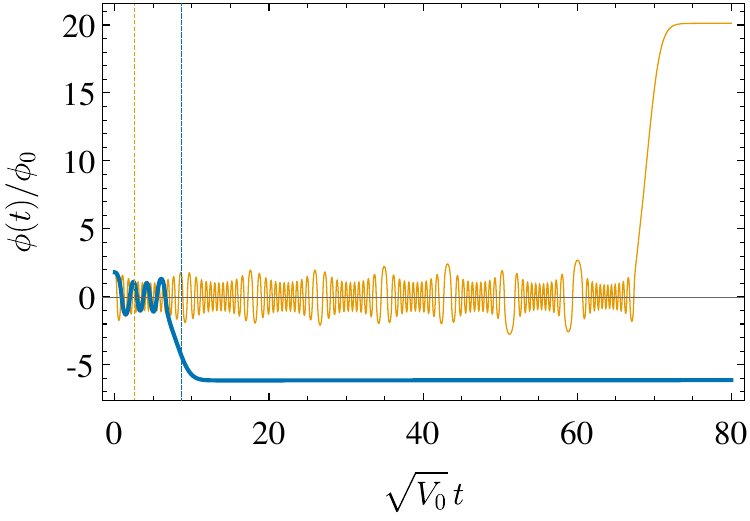} \par
    \vspace{1mm}
    \includegraphics[width=0.99\columnwidth]{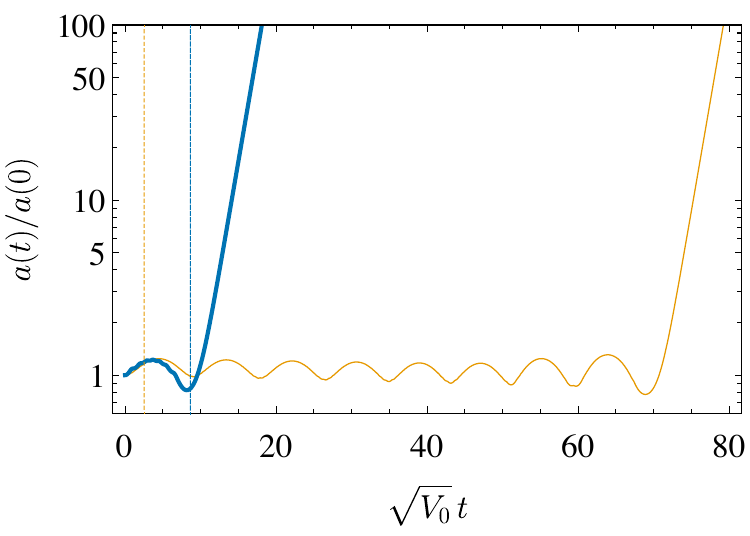}
    \caption{Two solutions representing the emergence of inflation after a quasi-cyclic period. The top panel shows $\phi(t)$ and the bottom panel shows the corresponding log plot of $a(t)$.  The solution shown by the \lineIdea{} lines ($\phi_0 = \sqrt{0.006}$ and $\Gamma/\sqrt{V_0} = 0.003 $) illustrates the beginning of inflation after the eight bounces, which is too late compared to the left vertical line showing $t_\text{frag}$ estimated by \CosmoLattice{} after fixing $V_0$ to fit $A_\text{s}$. The other solution shown by the \lineBenchMark{}  lines ($\phi_0 = \sqrt{0.06}$ and $\Gamma /\sqrt{V_0}= 0.02 $) illustrates one bounce followed by the beginning of inflation slightly before the right vertical line showing $t_\text{frag}$ estimated by \CosmoLattice{} after fixing $V_0$ to fit $A_\text{s}$.}
    \label{fig:dissipative_genesis}
\end{figure}

We use the set of equations of motion above to study the dynamics just after the creation of the Universe. We find that inflation longer than 60 e-folding is possible after undergoing single or multiple bounces depending on the parameters. Two such examples are shown in Fig.~\ref{fig:dissipative_genesis}. We have confirmed that qualitatively similar solutions are obtained even if we multiply $\Gamma$ with a ``window function'' to suppress  dissipation when the effective mass of $\phi$ is negative or small. 

The beginning of inflation due to the dissipation effects is not always guaranteed.  If the dissipation rate is too large or too small, the other generic consequence after the quasi-cyclic epoch is dominance either by the kinetic energy or by radiation and the resultant collapse of the Universe into the big crunch singularity.  In particular, once the Universe contracts so that the energy density of $\phi$ exceeds the asymptotic value $V_0$, there is no longer a chance for a bounce. 
This also suggests that sufficiently small shift-symmetry breaking terms can help the Universe to bounce. 
 Successful parameter values of $\Gamma$ are distributed dispersedly corresponding to different numbers of times of bounces.

These qualitative features do not sensitively depend on the details of the potential as long as it has a sufficiently flat part.  We have studied other example potentials and found similar solutions.  In particular, the plateau does not need to be infinitely extended.

So far, we have neglected the inhomogeneity of $\phi$, but there is an intrinsic tachyonic instability for any plateau-type potential. The tachyonic instability turns out to be so strong that the number of cycles is significantly limited. 
When the gradient energy becomes, e.g., $1 \%$ of the initial energy density, the backreaction to the background becomes non-negligible and $\phi$ will fragment. 
This fragmentation time is estimated as 
\begin{align}
    t_\text{frag} 
      \approx & \frac{1}{2 \mu_\text{peak}}  
      \ln \left( \frac{4\pi^{3/2} \times 0.01 
       \rho}{(k_\text{peak}/a)^4} \right) ,
\end{align}
where $\mu_\text{peak}$ is the maximum exponent of the tachyonic instability and $k_\text{peak}/a $ is the corresponding wave number~\cite{Tomberg:2021bll}, both of which are of $\mathcal{O}(\sqrt{2V_0}/\phi_0)$. The dimensionless fragmentation time $t_\text{frag}\sqrt{V_0}$ can be extended by lowering the energy scale $V_0$ but only logarithmically.

We estimate $t_\text{frag}$ more precisely using the public lattice calculation tool \CosmoLattice~\cite{Figueroa:2020rrl, Figueroa:2021yhd} (see the accompanying paper~\cite{Matsui:2023ezh} for details) where we turn off the dissipation effects, and the results are shown by the vertical dashed lines in Fig.~\ref{fig:dissipative_genesis}.
We have found a set of parameters consistent with the CMB observations with which inflation begins before the fragmentation time (the right vertical lines), which is shown by the \lineBenchMark{} lines, justifying the above calculations with homogeneity. It may appear that the inflation barely begins before $t = t_\text{frag}$ and one might doubt the validity of the solution.  At $t=t_\text{frag}$, however, $\phi$ has already reached deep on the plateau, so the fragmentation cannot occur thereafter and inflation safely begins.  On the other hand, the parameter values for the \lineIdea{} lines are chosen to clearly demonstrate the quasi-cyclic behavior and are incompatible either with the CMB normalization or with the bound from tachyonic instability (the left vertical lines). 

The tachyonic instability becomes more severe for smaller $\phi_0$ since $\phi$ oscillates more in the time scale of the dynamics of $a$. Meanwhile, for larger $\phi_0$, the cyclic solutions themselves (without dissipation effects) require more fine-tuned parameters.\footnote{For $\mathcal{O}(1)$ values of $\phi_0$ with fine-tuned initial conditions, there are solutions with inflation after a single or a few bounce(s) even without dissipation effects~\cite{Matsui:2019ygj, Sloan:2019jyl}. We see that similar solutions exist in an yet unexplored parameter space in the next section.}  Interestingly, there exists an allowed window for $\phi_0 \approx 0.25$, which predicts $r \approx 4 \times 10^{-5} \, (N_\mathrm{e}/55)^{-2}$. 
In this case, the number of cycles is limited to one or a few, at most.

\section{Exploring the parameter space}

Let us study where in the parameter space $(\phi_0, V_0, \phi(0), \Gamma)$ the mechanism works. The two most important parameters are the location of the plateau $\phi_0$ and the dissipation rate $\Gamma$.  The potential height $V_0$ rescales the overall timescale and affects the ratio between $t_\text{frag}$ and the typical time scale of the dynamics of $\phi$ logarithmically. We fix it to fit the CMB normalization.  The initial field value $\phi(0)$ is also an important factor that affects the behavior of solutions.  

\begin{figure}[htbp]
    \centering
    \includegraphics[width=0.93 \columnwidth]{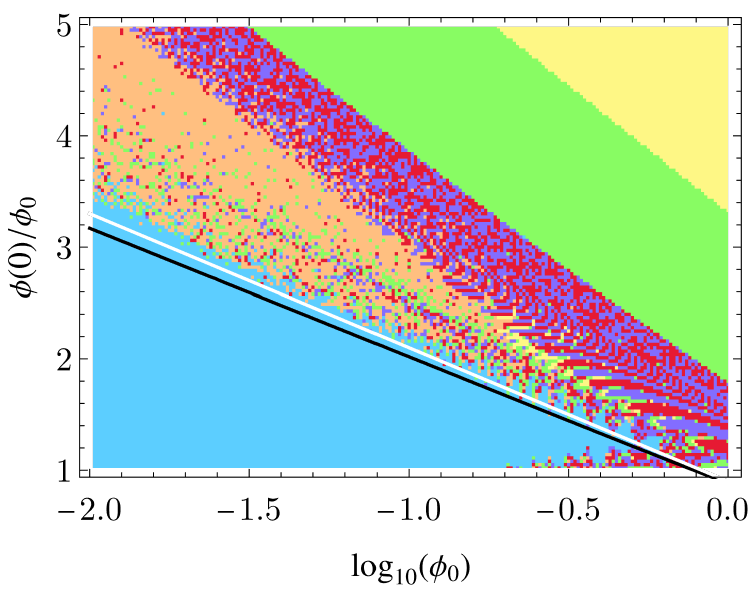} 
    \caption{Classification of the parameter points without dissipative effects. (1) \colorBigCrunchPositive, (2) \colorBigCrunchNegative, (3) \colorCyclic, (4) \colorShortInflation,  (5) \colorLongInflationTachyonic, and (6) \colorLongInflation{} regions denote big crunch with (1) $\dot{\phi} > 0$ or (2) $\dot{\phi} < 0$, (3) cyclic universes, (4) short ($\ln a(t_\text{max})/a(0) < 50$) inflation, and long ($\ln a(t_\text{max})/a(0) \geq 50$) inflation (5) after or (6) before the fragmentation time, respectively. The figure depends on $V_0$ only through the fragmentation time calculation for the region (5), and $V_0$ is fixed to fit the CMB normalization.  The slice of the parameter space shown by the \colorChoice{} line is further studied in Fig.~\ref{fig:phase_diagram}. The \colorFlatness{} line denotes the place where the first slow-roll parameter is $1$.}
    \label{fig:phase_diagram_basic}
\end{figure}

We first study where in the parameter space one finds cyclic solutions before considering  the effects of dissipation (i.e., $\Gamma = 0$).  Fig.~\ref{fig:phase_diagram_basic} classifies solutions with different $\phi_0$ and $\phi(0)$. We set the maximal time $t_\text{max} = 300/\sqrt{V_0}$ for the classification of solutions. 
In the upper-right corner, there are inflationary solutions with an e-folding number (within the simulation time) greater than $50$ (\colorLongInflation) or smaller than that (\colorShortInflation) as expected. Below that, there are parameter regions that end with a big crunch (\colorBigCrunchPositive{} for $\dot{\phi}>0$ and \colorBigCrunchNegative{} for $\dot{\phi}<0$), which is also not surprising.  There are also non-trivial regions in the parameter space.
The \colorCyclic{} region, located in the lower-left corner, represents cyclic solutions.  The existence of inflationary solutions (\colorLongInflation{} and \colorShortInflation) just above the \colorCyclic{} region is also remarkable. Unlike the inflationary solutions around the upper-right corner, inflation occurs after cosmic bounce(s) in this region. Thus, inflation after bounce(s) is possible even in the absence of dissipation. The \colorLongInflationTachyonic{} regions (passing through the top-left corner) represent sufficiently long inflation in the homogeneous simulation, but the beginning of the inflation is later than the fragmentation time.  Dissipative effects are also useful for such solutions to trigger inflation earlier. 

Next, we take into account the dissipative effects and study how the solutions are modified.  We expect solutions with larger amplitude $\phi(0)$ are more susceptible to dissipative effects. Also, the classicality condition $|V'|\lesssim V$ excludes insufficiently large $\phi(0)$ (see the \colorFlatness{} line as a guide below which the first slow-roll parameter is larger than $1$).  For these reasons, let us study a parameter region near the upper boundary of cyclic phase. For definiteness, we focus on the slice shown by the \colorChoice{} line in Fig.~\ref{fig:phase_diagram_basic}: $\phi(0)/\phi_0 = 0.9 - 1.2 \log_{10} \phi_0$.

\begin{figure}[htbp]
    \centering
    \includegraphics[width=0.99\columnwidth]{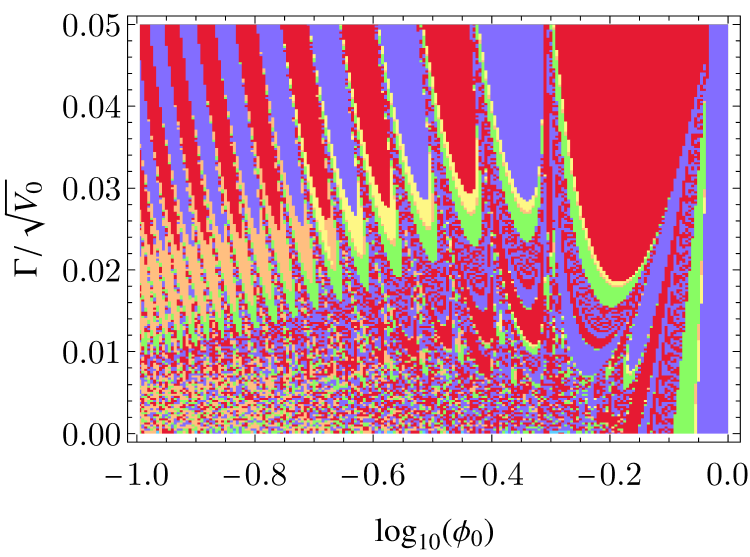} 
    \caption{Classification of the parameter points. (1) \colorBigCrunchPositive, (2) \colorBigCrunchNegative, (3) \colorCyclic, (4) \colorShortInflation, (5) \colorLongInflationTachyonic, and (6) \colorLongInflation{} regions denote big crunch with (1) $\dot{\phi} > 0$ or (2) $\dot{\phi} < 0$, (3) cyclic universes, (4) short inflation, and long ($\ln a(t_\text{max})/a(0) \geq 50$) inflation (5) in the absence of tachyonic instability or (6) even in its presence, respectively. 
    We choose the value of $V_0$ to fit the CMB normalization and that of $\phi(0)$ through the relation $\phi(0)/\phi_0= 0.9 - 1.2 \log_{10} \phi_0$ (the \colorChoice{} line in Fig.~\ref{fig:phase_diagram_basic}).}
    \label{fig:phase_diagram}
\end{figure}

The behavior of the solutions on the ($\phi_0$, $\Gamma$)-plane is summarized in Fig.~\ref{fig:phase_diagram} with a similar color coding. The characteristic stripe structures correspond to the oscillation phases of $\phi$ at which nontrivial dynamics occur.   The figure shows that there are 
 \colorLongInflation{} regions in the parameter space in which sufficiently long inflation occurs before the would-be fragmentation time. 
 In \colorLongInflationTachyonic{} regions, long inflation occurs in the homogeneous simulation, but it begins later than the fragmentation time.  If we remove the minimal assumption that this inflation is responsible for the CMB fluctuations, we can lower $V_0$ and increase the fragmentation time. Then, the \colorLongInflationTachyonic{} regions could become as viable as in the \colorLongInflation{} regions.

\section{Reheating of the Universe}
Once inflation occurs, the spatial curvature is flattened out and radiation is diluted, so the phenomenology is the same as the conventional inflationary scenario. An interesting point in our scenario is that the process of reheating the Universe is due to the same interactions that triggered inflation.  
 This gives us a chance to probe the pre-inflationary physics immediately after the creation of the Universe by studying the reheating. 
This is especially interesting if the Standard Model fields play an important role as the radiation $\rho_\text{r}$ in the above equations.  If the radiation consists of the dark/hidden sector beyond the Standard Model, we need to understand the coupling between such a sector and the Standard Model sector for a successful reheating.  However, this connection between the reheating and the quasi-cyclic period, the pre-/post-inflation duality, is a unique feature in our scenario. 

It is also worth noting that the dynamics of $\phi$ before and after inflation are quite different since the spatial curvature plays a major role in the former case whereas it is negligible in the latter case.  This difference can also be understood from the impact of the Hubble expansion/contraction rate.  Before inflation, it oscillates around zero, so the physics is closer to that in flat spacetime.  On the other hand, the Hubble friction significantly reduces the amplitude of $\phi$ after inflation, so preheating due to self-interactions is not effective for $\phi_0 \approx 0.25$~\cite{Lozanov:2017hjm}. We have confirmed this by lattice calculations using \CosmoLattice.

\section{Probability of inflation}

So far, we have treated $\phi(0)$ as a free parameter. 
Its probability distribution $\mathcal{P}(\phi(0))$ is determined by the wave function of the Universe and has the following exponential dependence~\cite{Vilenkin:1987kf}
\begin{align}
\mathcal{P}(\phi(0))\sim \exp \left( \pm \frac{24\pi^2}{V(\phi(0))} \right), \label{probability}
\end{align}
where the plus (minus) sign corresponds to the no-boundary (tunneling) proposal. The most probable outcomes crucially depend on the sign, and this ambiguity has been a long-standing controversy in quantum cosmology~\cite{Hartle:1983ai,Vilenkin:1986cy,Vilenkin:1987kf}.

Our study does not contribute to the theoretical resolution of the controversy, and we have nothing to say about the tunneling proposal since long-lasting inflation can easily occur.  
On the other hand, if the wave function of our Universe is close to that predicted by the no-boundary proposal, our solutions have a potentially significant impact. It has long been thought that the potential like Eq.~\eqref{V} most probably does not lead to long-lasting inflation for the no-boundary proposal~\cite{Vilenkin:1987kf} since a smaller $V(\phi(0))$ is more favored probabilistically. 
However, this statement must be reconsidered. 
Since the probability~\eqref{probability} is derived by assuming the flatness condition on the potential, let us exclude the field range of $\phi$ that does not satisfy the condition.  The probability formula then tells us that the initial value of $\phi$ is exponentially more probable to be at the edge of the plateau where the flatness condition is barely satisfied~\cite{Matsui:2020tyd}. 
A conventional expectation in such a case is that inflation is too short and the Universe is supposed to end with the big crunch singularity. Our solutions show that this fate can be avoided by transitioning into inflation.

Still, we cannot conclude that such nontrivial dynamics are the most probable outcome in our model setup once we introduce the observed small positive cosmological constant $\Lambda$~\cite{Planck:2018vyg, Brout:2022vxf}. In the no-boundary proposal, nucleation of a universe at the global minimum $V(\phi(0))=\Lambda$ is exponentially more probable.  If such a low-energy universe were created  directly from nothing, it would be an empty universe with no galaxies and no observers. Given the present Universe with galaxies and our existence, we may focus on the universes undergoing long-lasting inflation, which corresponds to thinking about the conditional probability.  Combining our solutions with this anthropic argument implies that the past of our inflationary Universe in the no-boundary proposal is most probably the quasi-cyclic period. 

\section{Conclusions}
We have found cosmological inflationary solutions in which the interactions that cause the reheating of the Universe after inflation are also responsible for dissipation to start inflation out of a quasi-cyclic epoch just after the quantum creation of the Universe from nothing.  One prediction of the scenario is positive spatial curvature. The connection between the pre- and post-inflationary periods can, in principle, help us experimentally and observationally probe the origin of our Universe.  

\section*{Acknowledgments}
Our analyses on the \CosmoLattice{} outputs were helped by the Mathematica notebook created by Francisco Torrenti that is available on \href{https://cosmolattice.net/}{the \CosmoLattice{} website}. 
T.T.~thanks the Yukawa Institute for Theoretical Physics (YITP) at Kyoto University. He thanks participants for discussions during the YITP workshop YITP-W-22-11 on ``Progress in Particle Physics 2022''.  
He also thanks participants of the workshop ``100+7 GR \& Beyond: Inflation'' for discussions. 
This work was supported by IBS under the project code, IBS-R018-D1, JSPS Core-to-Core Program (Grant No.~JPJSCCA20200002) (F.T.),  JSPS KAKENHI Grants No.~22KJ1782 (H.M.), No.~23K13100 (H.M.),
No.~20H01894 (F.T.), and No.~20H05851 (F.T.). This article is based upon work from COST Action COSMIC WISPers CA21106,  supported by COST (European Cooperation in Science and Technology).

\bibliographystyle{apsrev4-1}
\bibliography{letter_rev2.bib}

\end{document}